\documentclass[letterpaper,oneside,final,notitlepage,12pt,onecolumn]{article}
\usepackage{amssymb,amsmath,cite,epsfig}
\usepackage[top=2.5cm, bottom=2.5cm, left=2.5cm, right=2.5cm]{geometry}
\usepackage{amssymb}
\usepackage{amsfonts}
\usepackage{amsmath}
\usepackage[T1]{fontenc}
\usepackage[utf8]{inputenc}
\usepackage[english]{babel}
\usepackage{graphicx}

\input{tcilatex}
\begin{document}

\title{Holographic dark energy in Gauss-Bonnet gravity with Granda-Oliveros
cut-off}
\author{M. Koussour$^{1,\thanks{%
e-mail: pr.mouhssine@gmail.com}}$ H. Filali$^{2,\thanks{%
e-mail: houda.filali318@gmail.com}}$, S.H. Shekh$^{3,\thanks{%
e-mail: da\_salim@rediff.com}}$ and M. Bennai$^{1,2,\thanks{%
e-mail: mdbennai@yahoo.fr}}$ \\
$^{1}${\small Quantum Physics and Magnetism Team, LPMC,}\\
{\small Faculty of Science Ben M'sik, Casablanca Hassan II University,
Morocco.}\\
$^{2}${\small Lab of High Energy Physics, Modeling and Simulations,}\\
{\small Faculty of Science, University Mohammed V-Agdal, Rabat, Morocco.}\\
$^{3}${\small Department of Mathematics. S. P. M. Science and Gilani Arts
Commerce College,} \\
{\small Ghatanji, Yavatmal, Maharashtra-445301, India.}}
\maketitle

\begin{abstract}
In this paper, we have studied a model of holographic dark energy (HDE) with
a homogeneous and anisotropic Bianchi type-I Universe in the framework of
Gauss-Bonnet (GB) gravity or $f(G)$ gravity. To find an exact solution of
the field equations, we assume that the deceleration parameter varies with
cosmic time $\left( t\right) $. We plot some physical quantities and give
interpretations of the results obtained. Finally, we compared our model with
the $\Lambda CDM$ model by analyzing the Jerk parameter, and we examine the
equivalence between the HDE of the present work and the generalized HDE.

\end{abstract}

\section{Introduction}

According to recent astronomical observations \cite{ref1, ref2, ref3, ref4,
ref5, ref6, ref7, ref8, ref9, ref10, ref11}, our Universe has entered a
phase of accelerated expansion. This discovery leads to the presence of a
component of unknown nature, called Dark Energy (DE), which has negative
pressure and represents $68\%$ of the total density of the Universe, it
behaves like a repulsive gravity. Its nature remains unknown today. The
simple proposition for DE is the cosmological constant $\left(
\Lambda\right) $, which Einstein introduced into the field equations in
General Relativity (GR) in another context. This cosmological constant has
an equation of state (EoS) parameter $\omega=-1$ and is considered to be
very consistent with the observation data. Faced with the difficulties
linked to its theoretically predicted order of magnitude with respect to
that of the observed vacuum energy \cite{ref12}, other dynamic models of DE
have been proposed, such as quintessence \cite{ref13, ref14}, phantom \cite%
{ref15}, k-essence \cite{ref16}, tachyons \cite{ref17}, chaplygin gas \cite%
{ref18} and so forth. There is another class of dynamic DE models, in which
we do not need to introduce any other form of energy, this class is called
modified gravity theories, that is, an accelerated expansion can be caused
by a modification in action. In other words, GR is no longer valid on
cosmological scales. Such as $f\left( T\right) $ gravity, $f\left(
R,G\right) $ gravity, $f\left( R,T\right) $ gravity, $f\left(
R,T,R_{\mu\nu}T^{\mu\nu}\right) $ gravity and $f\left( T,T\right) $ gravity,
where $T$ is the trace of the energy-momentum tensor, $G$ is the
Gauss-Bonnet (GB) invariant and $R_{\mu\nu}$ is the Ricci tensor \cite%
{ref19, ref20, ref21, ref22}. The modified gravity of GB or the $f\left(
G\right) $ gravity, is considered among the modified versions of GR in which
we replace $R$ by $R+f\left( G\right) $ (where $f\left( G\right) $ is an
arbitrary function of the GB invariant $G$) in the Einstein-Hilbert action 
\cite{ref23, ref24}. Several authors have studied the applications of this
theory. The generalized second law of thermodynamics in cosmology in the
framework of the modified GB theory of gravity is investigated by Sadjadi 
\cite{ref25}. Myrzakulov et al. \cite{ref26} studied cosmological solutions,
especially the well-known $\Lambda CDM$ model. It is shown that the DE
contribution and even the inflationary epoch can be explained in the frame
of this kind of theory with no need for any other kind of component. The
cosmological application of holographic DE (HDE) in the framework of $%
f\left( G\right) $ modified gravity was discussed by Jawad et. al. \cite%
{ref27}. Although these works strongly inspire us to propose a new
cosmological model in the framework of $f\left( G\right) $ gravity and to
discuss the most important problems of the Standard Model of cosmology like
DE, there are other more important motivations, as this theory can describe
the current cosmic acceleration by passing the solar system tests for some
specific choices of $f\left( G\right) $ gravity models \cite{ref23}.

The holographic principle (HP) is another alternative to solve the problem
of DE, as it is known in the literature because it has great potential to
solve many long-standing problems in various physical fields. This principle
was first proposed by Hooft \cite{ref28} in the context of black hole
physics, then in a cosmological context another version of HP was proposed
by Fischler and Susskind \cite{ref29}. In the context of the DE problem, the
HP tells us that all physical quantities in the Universe, including the
density of DE $\left( \rho_{\Lambda}\right) $, can be described by a few
quantities on the boundary of the Universe. It is clear that in terms of two
physical quantities, namely the reduced Planck mass $\left( M_{p}\right) $\
and the cosmological length scale $\left( L\right) $, as follows $%
\rho_{\Lambda }\approx c^{2}M_{p}^{2}L^{-2}$ \cite{ref30}. Subsequently, a
relation was proposed which combines the HDE density $\left(
\rho_{\Lambda}\right) $ and the Hubble parameter $\left( H\right) $ as $%
\rho_{\Lambda}\propto H^{2}$, it does not contribute to the current
accelerated expansion of the Universe \cite{ref31}. For purely dimensional
reasons, Granda and Oliveros \cite{ref32} proposed a new infrared (IR)
cut-off for the holographic DE density of the form $\rho_{\Lambda}\approx%
\alpha H^{2}+\beta\overset{.}{H}$ where $\alpha$ and $\beta$ are constants.
They show that this new model of DE represents the accelerated expansion of
the Universe and is consistent with current observational data. Sarkar in
numerous works, studied the holographic model of DE in various contexts \cite%
{ref33, ref34, ref35}. Samanta studied the homogeneous and anisotropic
Bianchi type-V Universe filled with matter and HDE components, and
established a correspondence between the HDE and quintessence DE \cite{ref36}%
. Recently, locally rotationally symmetric (LRS) Bianchi type-I models with
HDE within the framework of $f(G)$ theory of gravitation are studied by
Shaikh et al. \cite{ref37}.

Recently, the anisotropic Universe has attracted the attention of many
researchers, because anisotropy played an important role in the early time
of cosmic evolution. In addition, the possibility of an anisotropy phase at
the beginning of the Universe followed by an isotropy phase was supported by
the observations i.e. the cosmic microwave background (CMB) anomalies from
the results obtained by Planck \cite{ref49}. Several researchers have
studied homogeneous and anisotropic Bianchi models, such as the spatially
homogeneous and anisotropic Bianchi type-I model, which is a direct
generalization of the FLRW Universe with a scale factor in each spatial
direction \cite{ref38, ref39}. In this paper, motivated by the work \cite%
{ref35}, we study the holographic model of DE under $f(G)$ gravity in the
Bianchi type-I Universe, to find solutions of the field equations and some
physical quantities, we will assume that the deceleration parameter (DP)
varies with time. The paper is organized as follows: Sect. 1 is an
introduction. In Sect. 2, we write the action of $f(G)$ gravity and the
field equation. We have derived the Bianchi type-I metric and defined some
physical and geometrical parameters to solve the field equations in Sect. 3.
In Sect. 4, we solve the field equations by considering the time-varying
deceleration parameter. Sect. 5 we discuss the jerk's parameter. Also, we
examine the equivalence between the HDE of the present work and the
generalized HDE in Sect. 6. The last section is devoted to a conclusion.

\section{Basic equations in Gauss-Bonnet gravity}

The modified Einstein--Hilbert action of the $f(G)$ gravity is given by \cite%
{ref23}

\begin{equation}
S=\frac{1}{2k^{2}}\int d^{4}x\sqrt{-g}\left[ R+f(G)\right] +S_{M}\left(
g^{\mu\nu},\psi\right) ,  \label{eqn1}
\end{equation}
where $k^{2}=8\pi G$, $R$ is the Ricci scalar, and $f\left( G\right) $ is a
general differentiable function of $GB$ invariant, $S_{M}$ is a matter
action that depends on a space-time metric $g_{\mu\nu}$ and matter fields $%
\psi$. The $GB$ invariant quantity is

\begin{equation}
G=R^{2}-4R_{\mu\nu}R^{\mu\nu}+R_{\mu\nu\alpha\beta}R^{\mu\nu\alpha\beta },
\label{eqn2}
\end{equation}
where $R_{\mu\nu}$ is the Ricci tensor and $R_{\mu\nu\alpha\beta}$ is the
Riemannian tensor.

The variation of the action (\ref{eqn1}) with respect to $g_{\mu\nu}$ leads
to the following equation

\begin{align}
& G_{\mu\nu}+8\left[ R_{\mu\rho\nu\sigma}+R_{\rho\nu}g_{\sigma\mu}-R_{\rho%
\sigma}g_{\nu\mu}-R_{\mu\nu}g_{\sigma\rho}+R_{\mu\sigma}g_{\nu\sigma }+\frac{%
1}{2}R\left( g_{\mu\nu}g_{\sigma\rho}-g_{\mu\sigma}g_{\nu\rho }\right) %
\right] \nabla^{\rho}\nabla^{\sigma}f_{G}  \label{eqn3} \\
& +\left( Gf_{G}-f\right) g_{\mu\nu}=k^{2}T_{\mu\nu},  \notag
\end{align}
where $\nabla_{\mu}$ denotes covariant differentiation, $G_{\mu\nu}=R_{\mu%
\nu }-\frac{1}{2}Rg_{\mu\nu}$ is the Einstein tensor, $T_{\mu\nu}$ is the
energy-momentum tensor of a matter fluid and the subscript $G$ in $f_{G}$
represents the derivative of $f$ with respect to $G$.

\section{Metric and field equations}

In this article, we will focus on a spatially homogeneous and anisotropic
Bianchi type-I Universe with different scale factors in each spatial
direction

\begin{equation}
ds^{2}=dt^{2}-A\left( t\right) ^{2}dx^{2}-B\left( t\right) ^{2}\left(
dy^{2}+dz^{2}\right) ,  \label{eqn4}
\end{equation}
where $A\left( t\right) $ and $B\left( t\right) $ are the directional metric
potentials.

The Ricci scalar and GB invariant for Bianchi type-I Universe are as follows

\begin{equation}
R=-2\left[ \frac{\overset{..}{A}}{A}+2\frac{\overset{..}{B}}{B}+2\frac{%
\overset{.}{A}\overset{.}{B}}{AB}+\frac{\overset{.}{B}^{2}}{B^{2}}\right] ,
\label{eqn5}
\end{equation}

\begin{equation}
G=8\left[ \frac{\overset{..}{A}\overset{.}{B}^{2}}{AB}+2\frac{\overset{.}{A}%
\overset{.}{B}\overset{..}{B}}{AB^{2}}\right] .  \label{eqn6}
\end{equation}

The energy-momentum tensor for matter and the HDE are defined as

\begin{equation}
\widetilde{T}_{\mu\nu}=\rho_{m}u_{\mu}u_{\nu},  \label{eqn7}
\end{equation}
and

\begin{equation}
\overline{T}_{\mu\nu}=\left( \rho_{\Lambda}+p_{\Lambda}\right)
u_{\mu}u_{\nu}+g_{\mu\nu}p_{\Lambda},  \label{eqn8}
\end{equation}
where $\rho_{m}$, $\rho_{\Lambda}$ are the energy densities of matter and
the HDE respectively and $p_{\Lambda}$ is the pressure of the HDE.

The field equations (\ref{eqn3}), with (\ref{eqn7}) and (\ref{eqn8}) for the
metric (\ref{eqn4}) leads to the following system of field equations

\begin{equation}
-2\frac{\overset{..}{B}}{B}-\frac{\overset{.}{B}^{2}}{B^{2}}+16\frac {%
\overset{.}{B}\overset{..}{B}}{B^{2}}\overset{.}{f}_{G}+8\frac{\overset{.}{B}%
^{2}}{B^{2}}\overset{..}{f}_{G}-Gf_{G}+f=k^{2}p_{\Lambda},  \label{eqn9}
\end{equation}

\begin{equation}
-\frac{\overset{..}{A}}{A}-\frac{\overset{..}{B}}{B}-\frac{\overset{.}{A}%
\overset{.}{B}}{AB}+8\left( \frac{\overset{.}{A}\overset{.}{B}}{AB}+\frac{%
\overset{..}{A}\overset{.}{B}}{AB}\right) \overset{.}{f}_{G}+8\frac{\overset{%
.}{A}\overset{.}{B}}{AB}\overset{..}{f}_{G}-Gf_{G}+f=k^{2}p_{\Lambda},
\label{eqn10}
\end{equation}

\begin{equation}
2\frac{\overset{.}{A}\overset{.}{B}}{AB}+\frac{\overset{.}{B}^{2}}{B^{2}}-24%
\frac{\overset{.}{A}\overset{.}{B}^{2}}{AB^{2}}\overset{.}{f}%
_{G}+Gf_{G}-f=k^{2}\left( \rho_{m}+\rho_{\Lambda}\right) ,  \label{eqn11}
\end{equation}
with an over dot $\left( \overset{.}{}\right) $ denote derivative with
respect to the cosmic time $\left( t\right) $.

The average scale factor $\left( a\right) $ of the Bianchi type-I Universe
is defined as

\begin{equation}
a=\left( AB^{2}\right) ^{\frac{1}{3}}.  \label{eqn12}
\end{equation}

The spatial volume $\left( V\right) $ of the Universe is given by

\begin{equation}
V=a^{3}=AB^{2}.  \label{eqn13}
\end{equation}

The average Hubble's parameter $\left( H\right) $ is defined as

\begin{equation}
H=\frac{\overset{.}{a}}{a}=\frac{1}{3}\left( H_{1}+2H_{2}\right) ,
\label{eqn14}
\end{equation}
where $H_{1}=\frac{\overset{.}{A}}{A}$ and $H_{2}=H_{3}=\frac{\overset{.}{B}%
}{B}$ are directional Hubble parameter along $x$, $y$ and $z$ axes
respectively.

For the Bianchi type-I Universe (\ref{eqn4}), the scalar expansion $\left(
\theta\right) $, deceleration parameter $\left( q\right) $\ and the shear
scalar $\left( \sigma^{2}\right) $ have the form

\begin{equation}
\theta=3H=\frac{\overset{.}{A}}{A}+2\frac{\overset{.}{B}}{B},  \label{eqn15}
\end{equation}

\begin{equation}
q=-\frac{a\overset{..}{a}}{\overset{.}{a}^{2}}=\frac{d}{dt}\left( \frac{1}{H}%
\right) -1,  \label{eqn16}
\end{equation}

\begin{equation}
\sigma^{2}=\frac{1}{2}\left[ \overset{3}{\underset{i=1}{\tsum }}H_{i}^{2}-%
\frac{1}{3}\theta^{2}\right] .  \label{eqn17}
\end{equation}

The average anisotropy parameter $\left( A_{m}\right) $ is defined by

\begin{equation}
A_{m}=\frac{1}{3}\overset{3}{\underset{i=1}{\tsum }}\left( \frac{\Delta H_{i}%
}{H}\right) ^{2}=6\left( \frac{\sigma}{\theta }\right) ^{2},  \label{eqn18}
\end{equation}
where $\Delta H_{i}=H_{i}-H$ and $H_{i}$ $\left( i=1,2,3\right) $ represent
the directional Hubble parameters.

By combining the HP and dimensional analysis, we find the HDE density as
follows \cite{ref30}

\begin{equation}
\rho_{HDE}=\frac{3C^{2}}{k^{2}L_{IR}^{2}},  \label{eqn19}
\end{equation}
where $C$ is a numerical constant that acts as a free parameter, $L_{IR}$ is
the infrared (IR) cut-off and $k^{2}=8\pi G$. In this work, we use a new
holographic Ricci dark energy model proposed by Granda and Oliveros \cite%
{ref32} for the HDE density

\begin{equation}
\rho_{\Lambda}=3\left( \alpha H^{2}+\beta\overset{.}{H}\right) ,
\label{eqn20}
\end{equation}
where $H$ is the average Hubble's parameter and $\alpha$, $\beta$ are
constants which must satisfy the constraints imposed by the current
observational data.

Combining (\ref{eqn9})--(\ref{eqn11}) the continuity equation can be
obtained as

\begin{equation}
\overset{.}{\rho}_{m}+\overset{.}{\rho}_{\Lambda}+3H\left(
\rho_{m}+\rho_{\Lambda}+p_{\Lambda}\right) =0.  \label{eqn21}
\end{equation}

The continuity equation of the matter is

\begin{equation}
\overset{.}{\rho}_{m}+3H\rho_{m}=0.  \label{eqn22}
\end{equation}

The continuity equation of the HDE is

\begin{equation}
\overset{.}{\rho}_{\Lambda}+3H\left( \rho_{\Lambda}+p_{\Lambda}\right) =0.
\label{eqn23}
\end{equation}

Using Eqs. (\ref{eqn20})--(\ref{eqn23}) and the barotropic equation of state 
$p_{\Lambda}=\omega_{\Lambda}\rho_{\Lambda}$, the equation of state HDE
parameter is obtained as

\begin{equation}
\omega_{\Lambda}=-1-\frac{2\alpha H\overset{.}{H}+\beta\overset{..}{H}}{%
3H\left( \alpha H^{2}+\beta\overset{.}{H}\right) }.  \label{eqn24}
\end{equation}

\section{Solutions of Field Equations}

The above field equations are nonlinear and complicated differential
equations, in order to solve these equations we assume that the deceleration
parameter varies with the time, which is of the form \cite{ref40}

\begin{equation}
q=-1+\gamma e^{-\gamma t},  \label{eqn25}
\end{equation}
where $\gamma>0$ is a constant.

Using the relation (\ref{eqn16}) and solving Eq. (\ref{eqn25}), we find

\begin{equation}
a=\left( e^{\gamma t}-1\right) ^{\frac{1}{\gamma}}.  \label{eqn26}
\end{equation}

Using the definition of the Hubble parameter we get

\begin{equation}
H=e^{\gamma t}\left( e^{\gamma t}-1\right) ^{-1}.  \label{eqn27}
\end{equation}

Using the relation $a(t)=\frac{1}{\left( 1+z\right) }$, with $z$ being the
redshift, gives us the following relation

\begin{equation}
t\left( z\right) =\frac{1}{\gamma}\log\left[ 1+\frac{1}{\left( 1+z\right)
^{\gamma}}\right] .  \label{eqn28}
\end{equation}

Also, the deceleration parameter $(q)$ can be written in terms of redshift $%
(z)$ as follows

\begin{equation}
q\left( z\right) =-1+\frac{\gamma\left( 1+z\right) ^{\gamma}}{1+\left(
1+z\right) ^{\gamma}}.  \label{eqn29}
\end{equation}

\begin{figure}[h]
\begin{center}
\includegraphics[height=8cm]{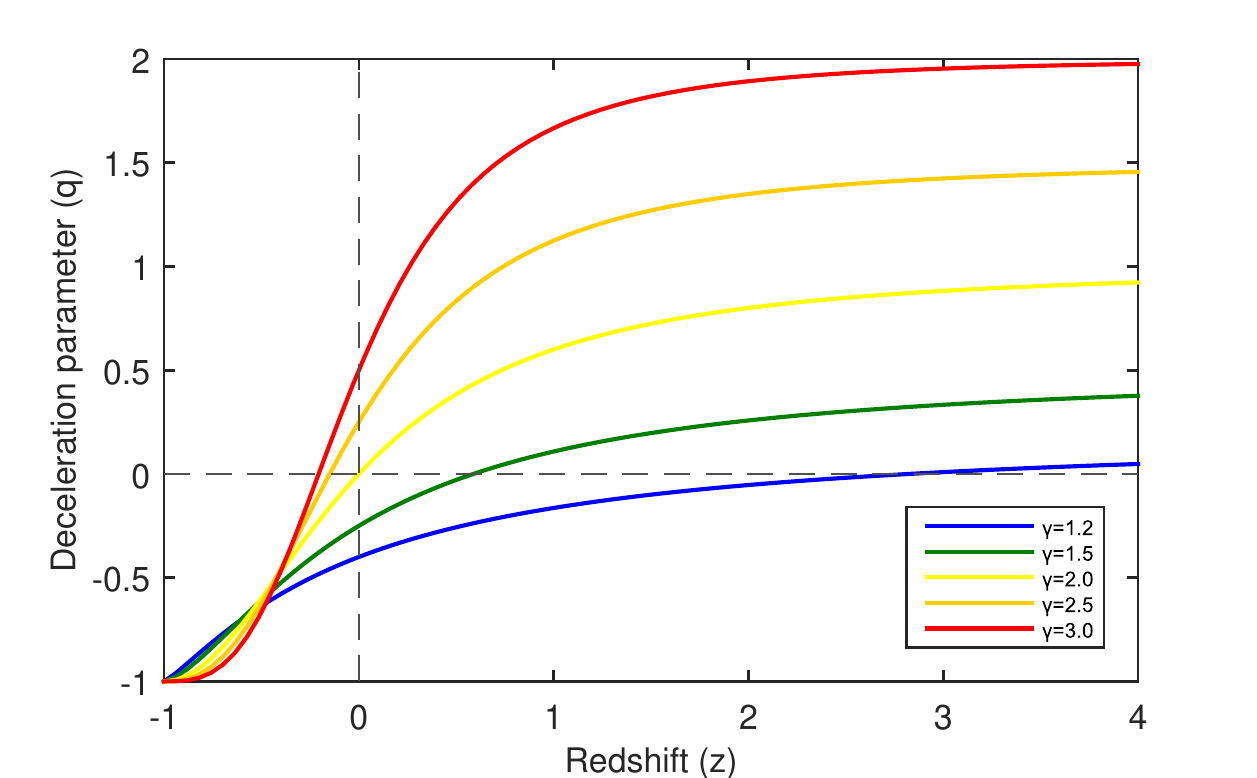}\newline
\end{center}
\caption{{\emph{Deceleration parameter versus redshift with $\gamma\geq1.2$.}%
}}%
\label{fig1}%
\end{figure}

The deceleration parameter is the quantity that describes the evolution of
the expansion of the Universe. This parameter is positive $\left( q>0\right) 
$ when the Universe is decelerated over time, and is negative $\left(
q<0\right) $\ otherwise, that is, when the expansion of the Universe is
accelerating. The current observational data \cite{ref41, ref42} indicates
that the Universe is accelerating and that the value of the deceleration
parameter is in the range $-1\leq q<0$. Fig. 1 shows the behavior of the
deceleration parameter in terms of redshift, it can be seen in Fig. 1 that
the deceleration parameter contains two phases in the Universe, the initial
deceleration phase and the current acceleration phase. In this work, to
produce both phases, we need $\gamma\geq1.2$. Also, the transition from the
early deceleration phase to the current accelerated phase is done with a
certain redshift, called a transition redshift $z_{tr}$. From the figure,
the value of the transition redshift for $\gamma=1.5$ is $z_{tr}=0.62$. This
transition redshift value is therefore consistent with the results of the
observation \cite{ref43, ref44, ref45}.

Similarly, we get the Hubble parameter $\left( H\right) $ in terms of the
redshift $\left( z\right) $ as

\begin{equation}
H\left( z\right) =1+\left( 1+z\right) ^{\gamma}.  \label{eqn30}
\end{equation}

The evolution of the Hubble parameter in terms of redshift with different $%
\gamma$ (i.e. $\gamma\geq1.2)$\ are shown in Fig. 2. It appears from this
figure that the Hubble parameter is a positive function in terms of
redshift. At present $\left( z=0\right) $ and early $\left( z>0\right) $,
the Hubble parameter is strictly positive and increases with increasing $z$
value. Also, the present HDE model does not admit any turning point in $%
H\left( z\right) $. Hence, our model has the same behavior in Ref. \cite%
{ref46}.

\begin{figure}[h]
\begin{center}
\includegraphics[height=8cm]{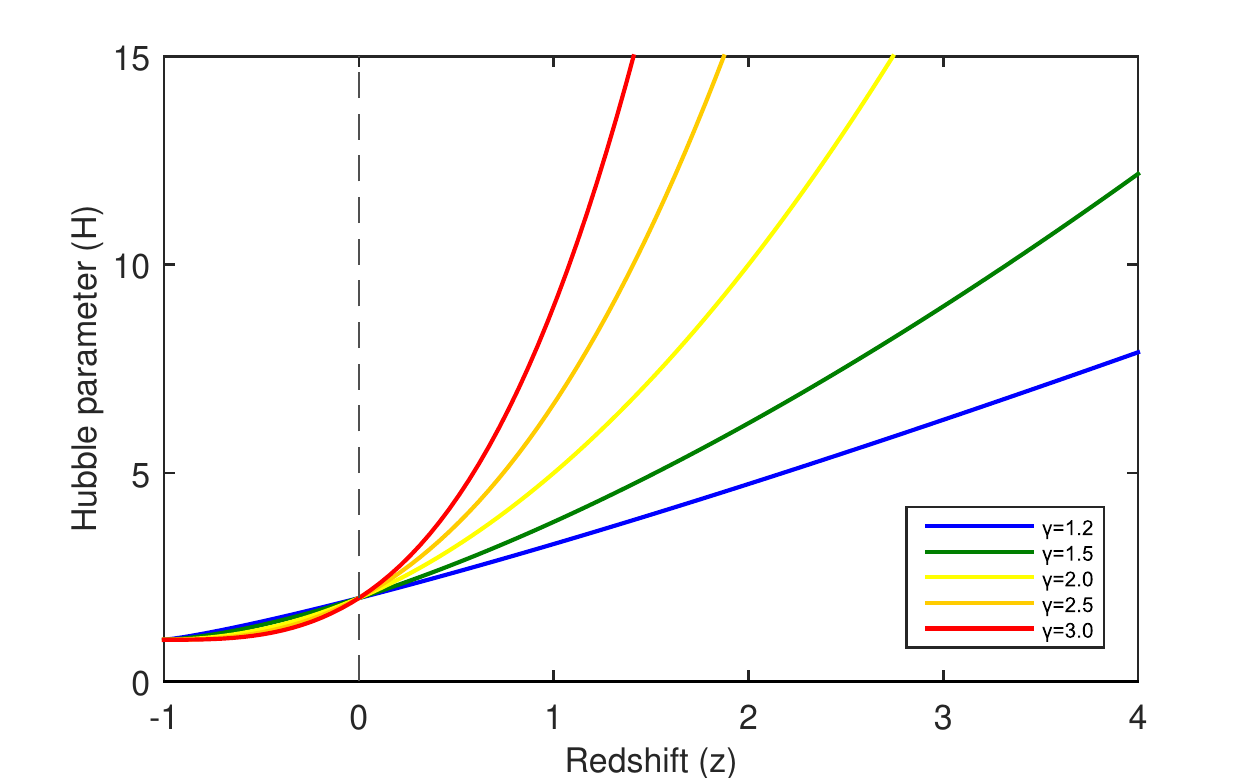}\newline
\end{center}
\caption{{\emph{Hubble parameter versus redshift with $\gamma\geq1.2$.}}}%
\label{fig2}%
\end{figure}

From the two Eqs. (\ref{eqn13}) and (\ref{eqn26}) above, the values of the
metric poentials $A$ and $B$ are obtained as

\begin{equation}
A=\left( e^{\gamma t}-1\right) ^{\frac{2}{\gamma}},  \label{eqn31}
\end{equation}

\begin{equation}
B=\left( e^{\gamma t}-1\right) ^{\frac{1}{2\gamma}}.  \label{eqn32}
\end{equation}

Here, we have used a condition so that the shear scalar is proportional to
the expansion scalar $\left( \theta\propto\sigma^{2}\right) $, and from it
we can write a relationship that sums the metric potentials as follows: $%
A=B^{m}$, where $m\neq1$ is an arbitrary constant. The model suit an
isotropic model for $m=1$, or else it suits anisotropic. In this paper we
have taken $m=4$, as a result, we get the metric potentials in Eqs. (\ref%
{eqn30}) and (\ref{eqn31}). The main reasons behind the assumptions that led
to this condition are discussed in detail here \cite{ref54}. Observations of
the velocity redshift relation for extragalactic sources indicate that the
Hubble expansion of the Universe can reach isotropic when $\frac{\sigma^{2}}{%
\theta}$ is constant. The condition has been used in many studies, see \cite%
{ref33, ref34, ref35, ref36}

Using Eqs. (\ref{eqn31}) and (\ref{eqn32}), the Bianchi type-I Universe
takes the form

\begin{equation}
ds^{2}=dt^{2}-\left( e^{\gamma t}-1\right) ^{\frac{4}{\gamma}}dx^{2}-\left(
e^{\gamma t}-1\right) ^{\frac{1}{\gamma}}\left( dy^{2}+dz^{2}\right) .
\label{eqn33}
\end{equation}

We suppose that the model of the function $f(G)$ follows the following
power-law models proposed by \cite{ref47}

\begin{equation}
f(G)=\eta G^{n+1},  \label{eqn34}
\end{equation}
where $\eta$ and $n$ are arbitrary constants. The motivations behind this
choice are numerous as mentioned by Shaikh et al. \cite{ref37}. For example,
the chances of the emergence of the Big-Rip singularity disappear, and also
the prediction of the existence of the transient phantom era consistent with
astrophysical observations are considered among the attracting factors in
the power-law model of $f(G)$ gravity. Some authors have worked on the
power-law of $f(G)$, such as \cite{ref48, ref49}.

The scalar expansion $\left( \theta\right) $, shear scalar $\left(
\sigma^{2}\right) $\ and the average anisotropy parameter $\left(
A_{m}\right) $\ are therefore obtained as

\begin{equation}
\theta=3e^{\gamma t}\left( e^{\gamma t}-1\right) ^{-1},  \label{eqn35}
\end{equation}

\begin{equation}
\sigma^{2}=\frac{3}{4}e^{2\gamma t}\left( e^{\gamma t}-1\right) ^{-2},
\label{eqn36}
\end{equation}

\begin{equation}
A_{m}=\frac{1}{2}.  \label{eqn37}
\end{equation}

From Eqs. (\ref{eqn35}) and (\ref{eqn36}) it appears that the scalar
expansion and the shear scalar diverge at $t\rightarrow0$ and they become
finite when $t\rightarrow\infty$. However, from Eq. (\ref{eqn37}) it is
observed that the anisotropic parameter remains constant throughout cosmic
evolution, which indicates that our model is anisotropic from the initial
era of the Universe to the final era.

Using Eq. (\ref{eqn27}) in (\ref{eqn20}), we get the HDE density as

\begin{equation}
\rho_{\Lambda}=3e^{\gamma t}\left( \alpha e^{\gamma t}-\beta\gamma\right)
\left( e^{\gamma t}-1\right) ^{-2}.  \label{eqn38}
\end{equation}

Again, using Eq. (\ref{eqn27}) in (\ref{eqn22}), we get the matter energy
density as

\begin{equation}
\rho_{m}=c_{1}\left( e^{\gamma t}-1\right) ^{-\frac{3}{\gamma}},
\label{eqn39}
\end{equation}
with $c_{1}$ is a constant of integration.

The coincidence parameter $\left( r\right) $\ can be defined as the ratio
between the HDE density $\left( \rho_{\Lambda}\right) $ and the matter
energy density $(\rho_{m})$, therefore for Eqs. (\ref{eqn38}) and (\ref%
{eqn39}) the coincidence parameter becomes

\begin{equation}
r=\frac{\rho_{\Lambda}}{\rho_{m}}=\frac{3}{c_{1}}e^{\gamma t}\left( \alpha
e^{\gamma t}-\beta\gamma\right) \left( e^{\gamma t}-1\right) ^{\frac {1}{%
\gamma}\left( 3-2\gamma\right) }.  \label{eqn40}
\end{equation}

From Fig. 3, it is shown that the HDE density $(\rho_{\Lambda})$ is a
decreasing function of cosmic time and positive throughout the evolution of
the Universe, its value being larger at the initial epoch and then disappear
later, which leads to the Universe ruled by a vacuum. The evolution of the
matter energy density $(\rho_{m})$ is illustrated in Fig. 4, where it starts
at a positive value, but also disappears later, which represents the
expansion of the Universe. The energy density of the Universe has not been
the same since the Big Bang and the beginning of the expansion of the
Universe. When the energy density is large, the radiation is more diffuse.
This period is called the radiation-dominated era and when the density is
low, it is the energy of the vacuum that dominates the Universe. Also, the
coincidence parameter $\left( r\right) $ as a function of cosmic time, it
turns out that the latter is an increasing function of cosmic time.
Therefore, at the beginning time, the Universe is dominated by HDE and later
the Universe is dominated by matter. This result is consistent with the
current Universe.

\begin{figure}[h]
\begin{center}
\includegraphics[height=8cm]{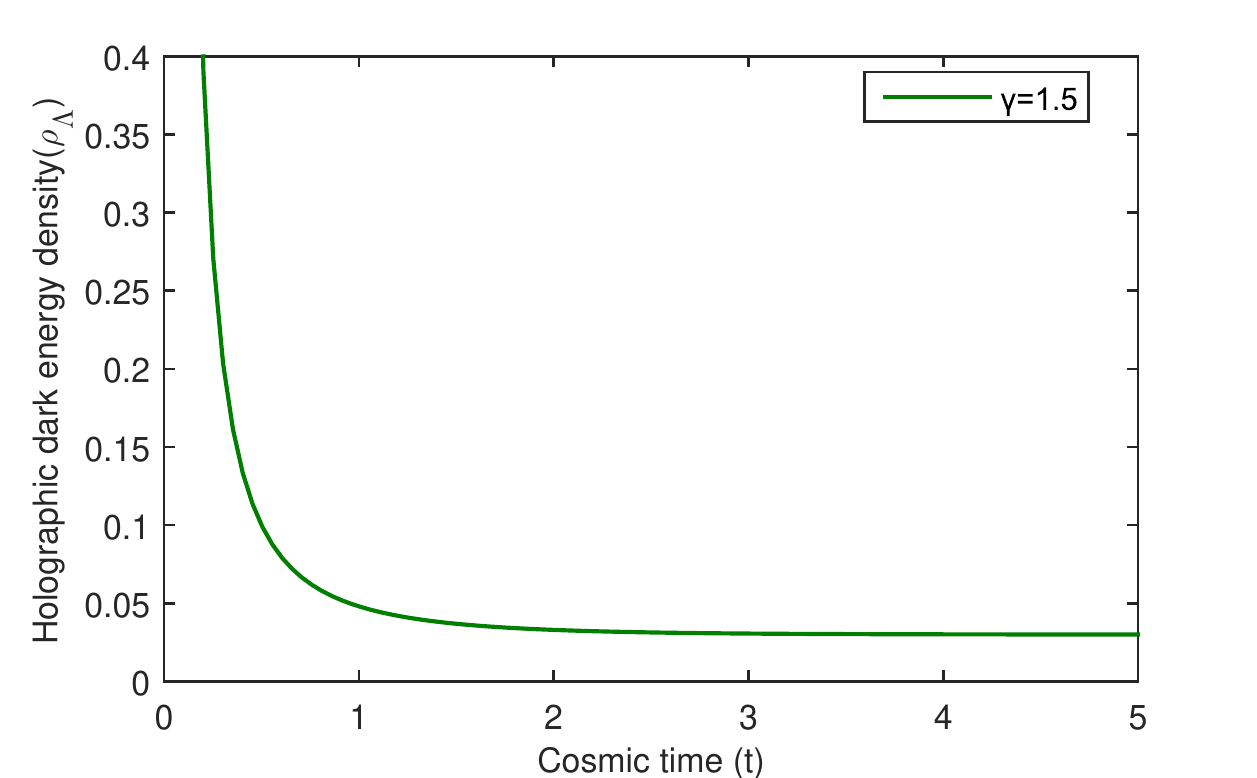}\newline
\end{center}
\caption{{\emph{An evolution of HDE density versus cosmic time with
$\alpha=0.01$, and $\beta=0.001$.}}}%
\label{fig3}%
\end{figure}

Now, using Eq. (\ref{eqn27}) in (\ref{eqn24}), we get the DE equation of
state parameter as

\begin{equation}
\omega_{\Lambda}=-1-\frac{\gamma\left[ e^{\gamma t}\left( \beta
\gamma-2\alpha\right) +\beta\gamma\right] }{e^{\gamma t}\left( 3\alpha
e^{\gamma t}-3\beta\gamma\right) }.  \label{eqn41}
\end{equation}

\begin{figure}[h]
\begin{center}
\includegraphics[height=8cm]{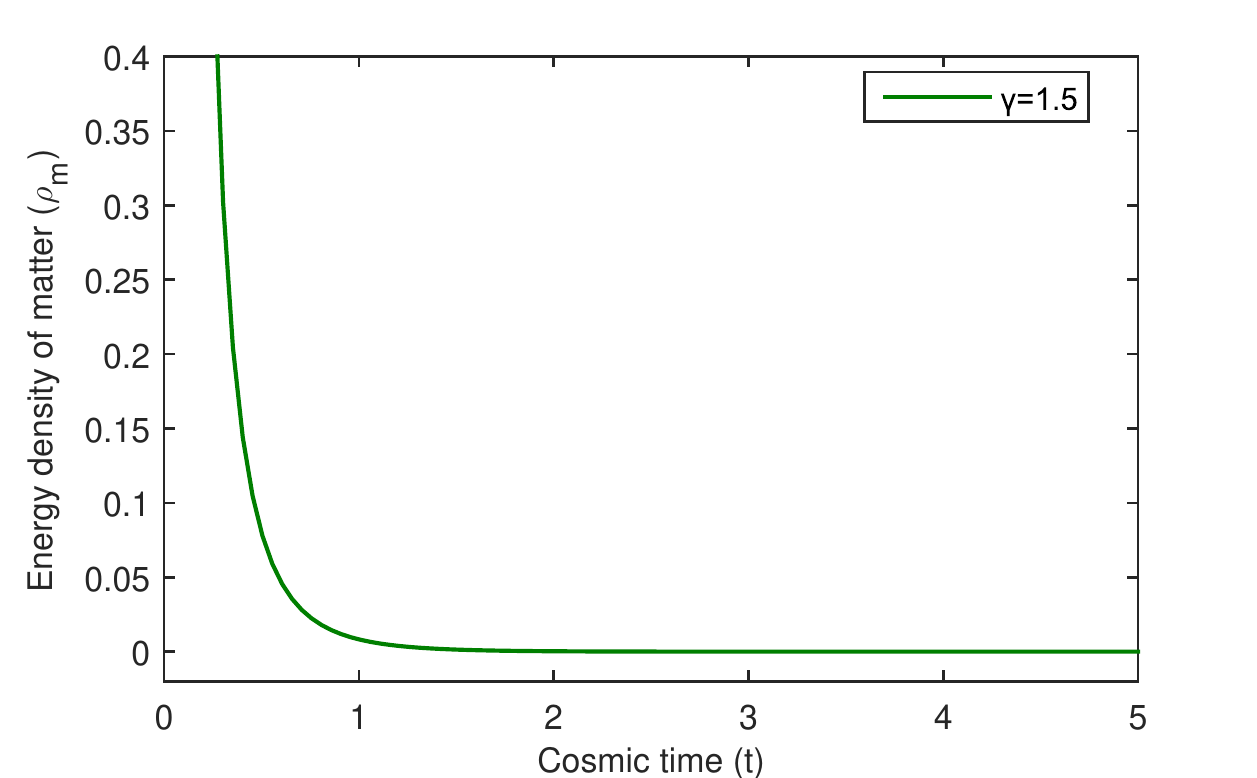}\newline
\end{center}
\caption{{\emph{An evolution of energy density of matter versus cosmic time
with $c_{1}=0.1$.}}}%
\label{fig4}%
\end{figure}

Fig. 5 clearly shows that the equation of state parameter evolves with
negative values in an appropriate range $\left(
-1\leq\omega_{\Lambda}\leq0\right) $, which is in good agreement with
astronomical observations. Our studied model is therefore realistic. From
Fig. 5, we notice that at the beginning of cosmic time the equation of state
parameter starts close to zero (that is to say the Universe dominated by
matter) and then at the end of cosmic time it takes a close negative value
of $-1$ (i.e. when the Universe dominated by the HDE). If $%
\omega_{\Lambda}=-1$, it represents the $\Lambda CDM$ model, $%
-1<\omega_{\Lambda}<-1/3$, represents the quintessence model and $%
\omega_{\Lambda}<-1$ indicates the phantom behavior of the model. Also, we
can observe that in the early Universe $-1<\omega_{\Lambda}<0$, it indicates
the quintessential model and in the current Universe, $\omega_{\Lambda}$
tends to $-1$, i.e. the model $\Lambda CDM$. We conclude from Fig. 6 that
the current value of the equation of state parameter of our model is in
rough agreement with recent observational data from Planck + WMAP \cite%
{ref50}.

The HDE pressure is obtained as

\begin{equation}
p_{\Lambda}=-\left[ e^{\gamma t}\left( 3\alpha e^{\gamma t}+\beta\gamma
^{2}-2\alpha\gamma-3\beta\gamma\right) +\beta\gamma^{2}\right] \left(
e^{\gamma t}-1\right) ^{-2}.  \label{eqn42}
\end{equation}

It is useful to use yet another notation, the abundances, also called the
density parameters, it represents the proportion of each element in the
Universe. The total energy density parameter $\left(
\Omega=\Omega_{m}+\Omega_{\Lambda}\right) $ takes three values $\Omega>1$, $%
\Omega=1$, $\Omega<1$ correspond respectively to the open, flat and closed
Universe. The matter density parameter $\left( \Omega_{m}\right) $\ and HDE
density parameter $\left( \Omega_{\Lambda}\right) $\ are respectively given
by

\begin{equation}
\Omega_{m}=\frac{\rho_{m}}{3H^{2}}=\frac{1}{3}c_{1}e^{-2\gamma t}\left(
e^{\gamma t}-1\right) ^{\frac{1}{\gamma}\left( 2\gamma-3\right) },
\label{eqn43}
\end{equation}
and

\begin{equation}
\Omega_{\Lambda}=\frac{\rho_{\Lambda}}{3H^{2}}=\alpha-\beta\gamma e^{-\gamma
t}.  \label{eqn44}
\end{equation}

\begin{figure}[h]
\begin{center}
\includegraphics[height=8cm]{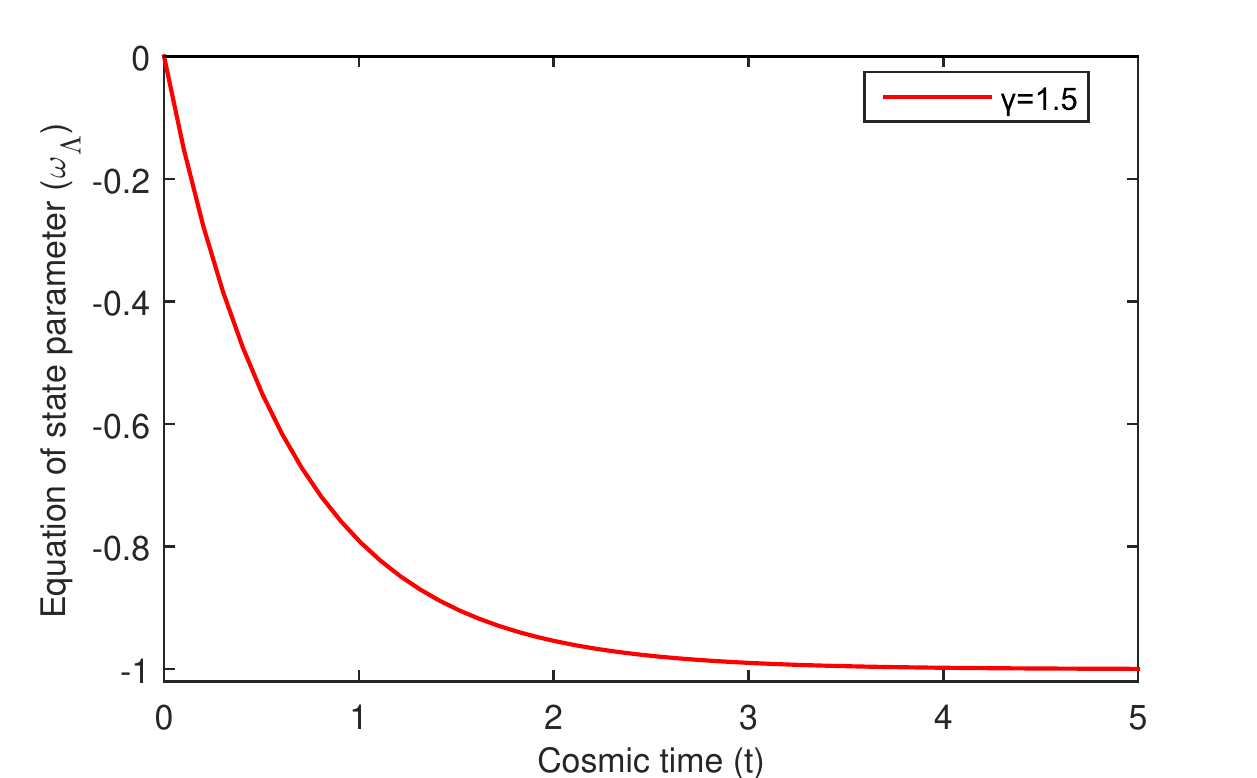}\newline
\end{center}
\caption{{\emph{An evolution of equation of state parameter versus cosmic time
with $\alpha=0.01$ and $\beta=0.001$.}}}%
\label{fig5}%
\end{figure}

\begin{figure}[h]
\begin{center}
\includegraphics[height=8cm]{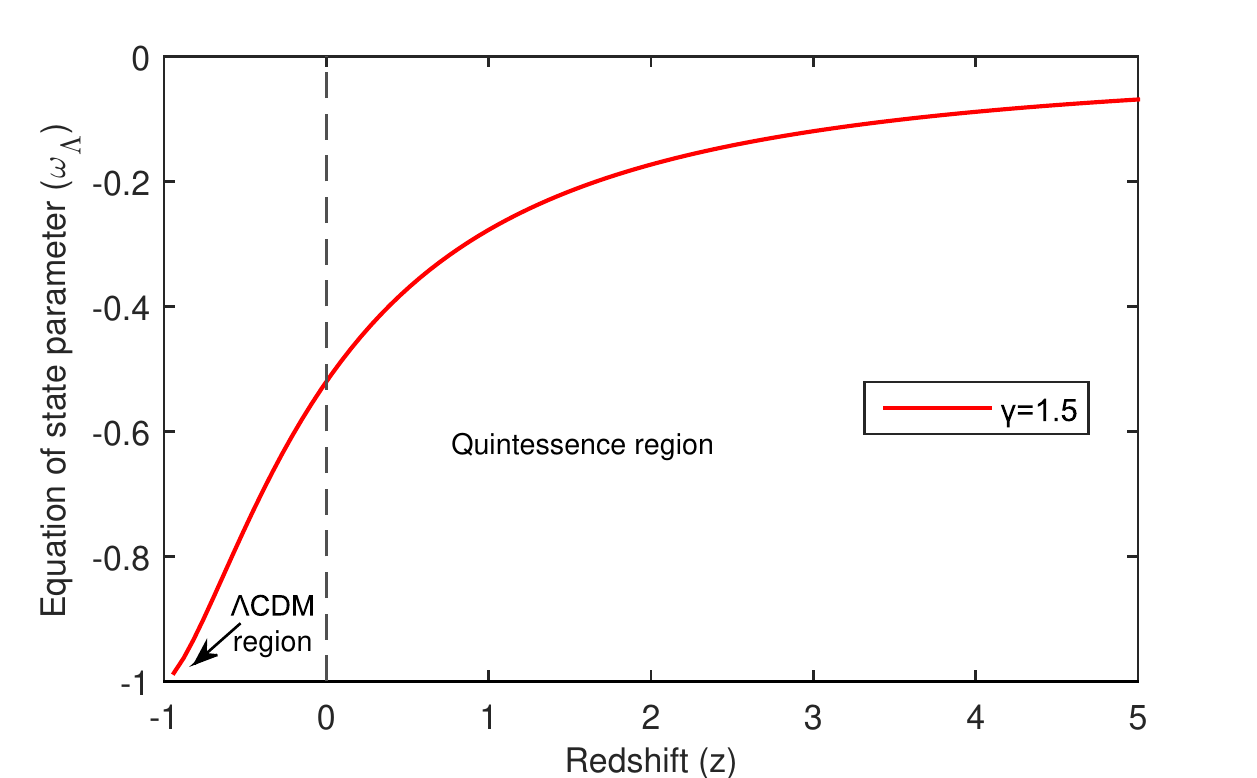}\newline
\end{center}
\caption{{\emph{An evolution of equation of state parameter versus redshift
with $\alpha=0.01$, and $\beta=0.001$.}}}%
\label{fig6}%
\end{figure}

We obtain the total density parameter yields as

\begin{equation}
\Omega=\frac{1}{3}c_{1}e^{-2\gamma t}\left( e^{\gamma t}-1\right) ^{\frac {1%
}{\gamma}\left( 2\gamma-3\right) }+\alpha-\beta\gamma e^{-\gamma t}.
\label{eqn45}
\end{equation}

Fig. 7 represents the evolution of the energy density parameter as a
function of time, and it appears that its value is large in the initial era
of the Universe, but we begin to approach $\Omega\sim1$ in the last era of
Universe, which causes our model to predict a flat Universe at a later time,
as recent astronomical observations indicate.

\begin{figure}[h]
\begin{center}
\includegraphics[height=8cm]{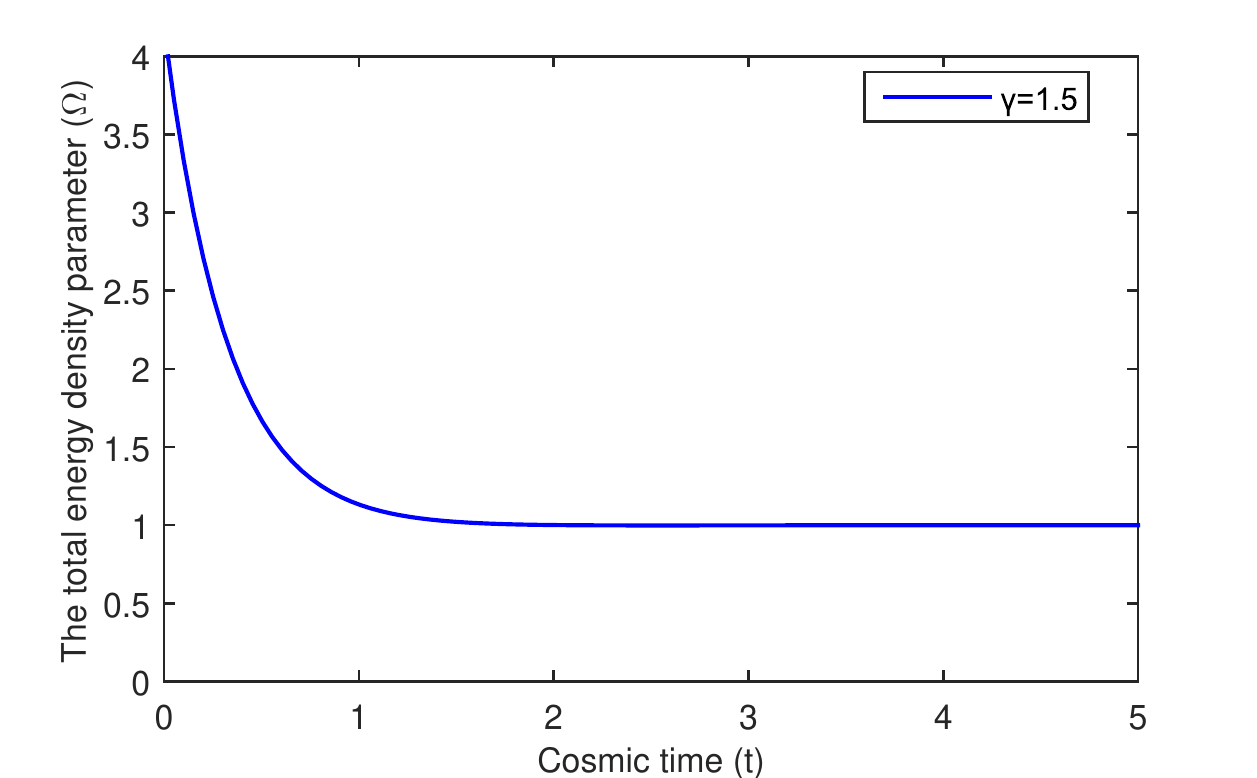}\newline
\end{center}
\caption{{\emph{The total energy density parameter $\left(  \Omega=\Omega
_{m}+\Omega_{\Lambda}\right)  $ versus cosmic time with $\alpha=0.01$,
$\beta=0.001$, and $c=0.1$.}}}%
\label{fig7}%
\end{figure}

We should note that under the assumption (\ref{eqn26}), the GB invariant $G$
and Ricci scalar $R$ behave as

\begin{equation}
G=12\left[ \left( 1-\gamma\right) e^{4\gamma t}\left( e^{\gamma t}-1\right)
^{-4}+\gamma e^{3\gamma t}\left( e^{\gamma t}-1\right) ^{-3}\right] ,
\label{eqn46}
\end{equation}

\begin{equation}
R=\left( 6\gamma-\frac{27}{2}\right) e^{2\gamma t}\left( e^{\gamma
t}-1\right) ^{-2}-6\gamma e^{\gamma t}\left( e^{\gamma t}-1\right) ^{-1}.
\label{eqn47}
\end{equation}

Using Eqs. (\ref{eqn34}) and (\ref{eqn46}), the function $f(G)$ is obtained
as

\begin{equation}
f\left( G\right) =\eta\left[ 12\left( 1-\gamma\right) e^{4\gamma t}\left(
e^{\gamma t}-1\right) ^{-4}+12\gamma e^{3\gamma t}\left( e^{\gamma
t}-1\right) ^{-3}\right] ^{n+1}.  \label{eqn48}
\end{equation}

\begin{figure}[h]
\begin{center}
\includegraphics[height=8cm]{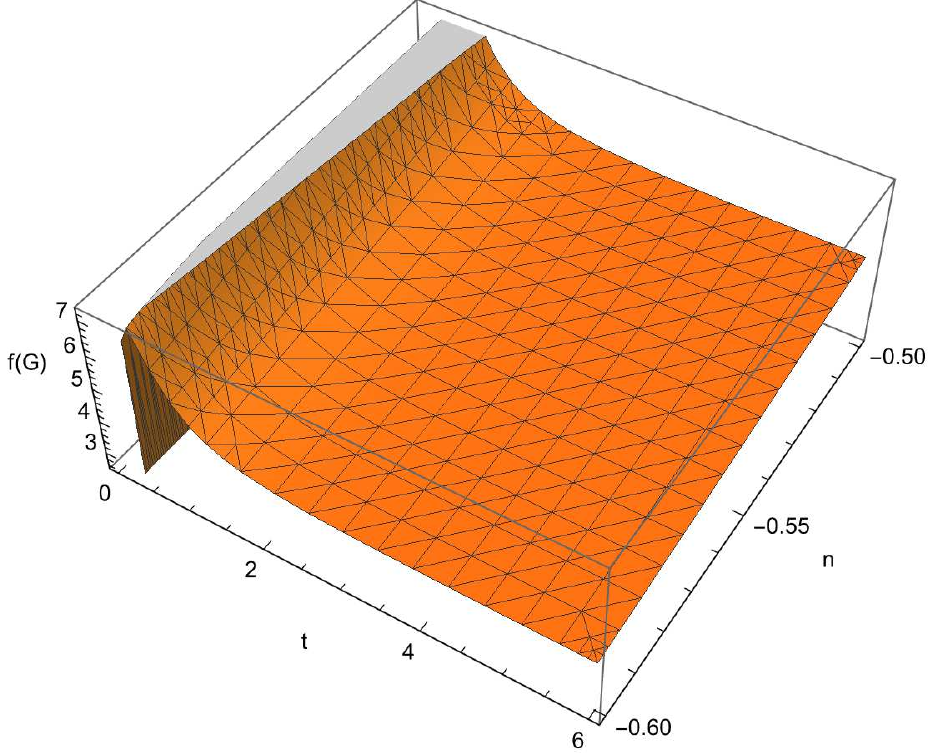}\newline
\end{center}
\caption{{\emph{An evolution of $f(G)$ model versus cosmic time and $n$ for
$\eta=\gamma=1.5$.}}}%
\label{fig8}%
\end{figure}

Fig. 8 represents the function $f(G)$ in terms of time for a range of values 
$n<0$. It shows that the function $f(G)$ is positive over cosmic time and
contains a transitory behavior. At the beginning of time, the function $f(G)$
starts with large values, then approaches zero, then increases, and finally
takes a constant value.

\section{The jerk parameter}

Among the simple tools to find deviations from the $\Lambda CDM$ concord
model, there is the Jerk parameter which is an important quantity to
describe the dynamic evolution of the Universe. We know that the models
close to the $\Lambda CDM$ model can be described by the jerk parameter, for
example for the flat $\Lambda CDM$ model the value of this parameter is
constant $j=1$. A deceleration to acceleration transition occurs for models
with a positive value of $j_{0}$ and negative $q_{0}$. The jerk parameter is
a third dimensionless derivative of the scale factor with respect to cosmic
time, in cosmology is defined as \cite{ref51, ref52, ref53}

\begin{equation}
j\left( t\right) =\frac{1}{H^{3}}\frac{\overset{..}{a}}{a}=q+2q^{2}-\frac{%
\overset{.}{q}}{H}.  \label{eqn49}
\end{equation}

For our model, the jerk parameter is given as follows

\begin{equation}
j\left( t\right) =1-3\gamma e^{-\gamma t}+2\gamma^{2}e^{-2\gamma
t}+\gamma^{2}e^{-2\gamma t}\left( e^{\gamma t}-1\right) .  \label{eqn50}
\end{equation}

Fig. 9 represents the evolution of the jerk parameter in terms of cosmic
time and shows that this parameter is positive throughout the evolution of
the Universe and for a large cosmic time, the jerk parameter tends towards $%
1 $, it approaches the $\Lambda CDM$ model.

\begin{figure}[h]
\begin{center}
\includegraphics[height=8cm]{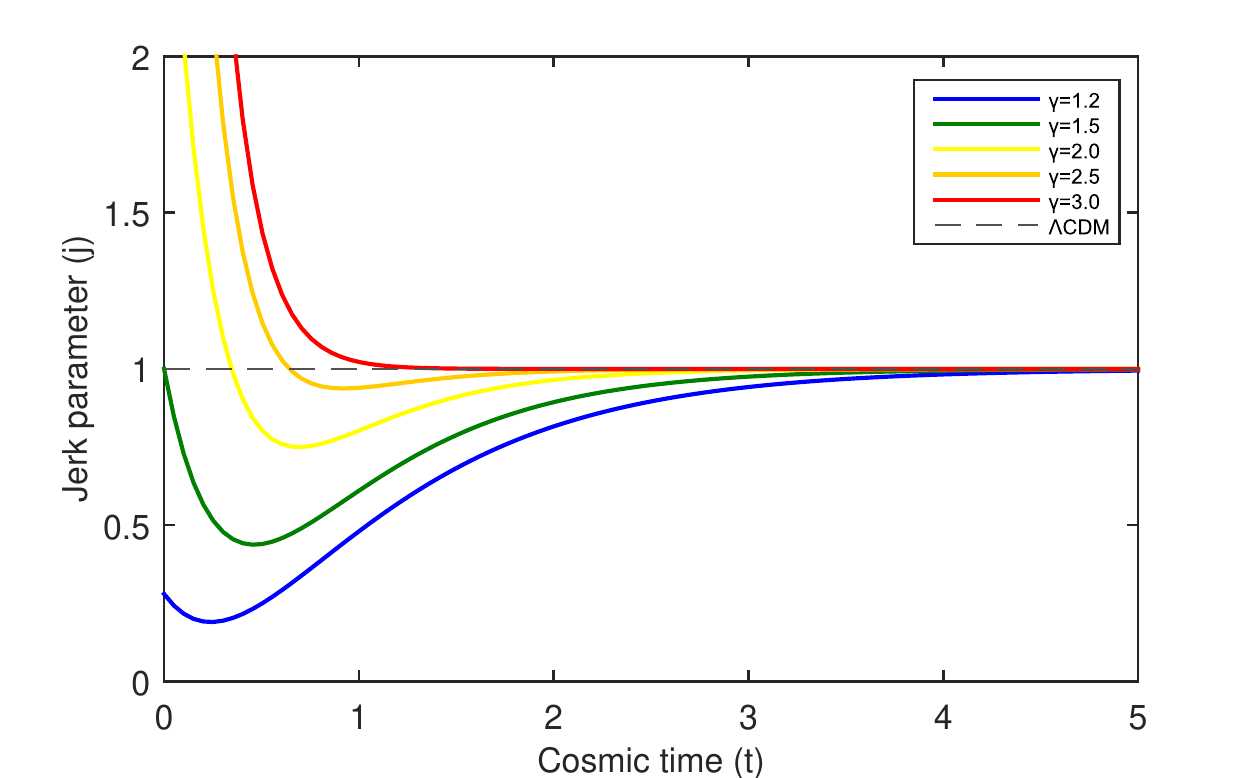}\newline
\end{center}
\caption{{\emph{9 An evolution of Jerk parameter versus cosmic time.}}}%
\label{fig9}%
\end{figure}

\section{Generalized Holographic Dark Energy}

In this section, we effort to establish that our dark energy model has a
direct equivalence to the generalized holographic dark energy model. As
shown in Eq. (\ref{eqn19}), the holographic dark energy density is inversely
proportional to the squared infrared cut-off $L_{IR}$. The IR cut-off is
supposed to be the particle horizon $L_{P}$ or the future event horizon $%
L_{F}$, which are determined respectively as \cite{ref55}

\begin{equation}
L_{P}\equiv a\int_{0}^{t}\frac{dt}{a},\text{ \ \ \ \ \ \ \ \ \ }L_{F}\equiv
a\int_{t}^{\infty}\frac{dt}{a}.  \label{eqn51}
\end{equation}

Adopted the time derivative of the above equation, we get the expressions
for the Hubble parameter in terms of $L_{P}$, $L_{F}$, and their time
derivatives as follows

\begin{equation}
H\left( L_{P},\overset{.}{L}_{P}\right) \equiv\frac{\overset{.}{L}_{P}}{L_{P}%
}-\frac{1}{L_{P}},\text{ \ \ \ \ \ \ \ \ \ }H\left( L_{F},\overset{.}{L}%
_{F}\right) \equiv\frac{\overset{.}{L}_{F}}{L_{F}}+\frac{1}{L_{F}}.
\label{eqn52}
\end{equation}

The general form of the cut-off is suggested in this work \cite{ref56}

\begin{equation}
L_{IR}=L_{IR}\left( L_{P},\overset{.}{L}_{P},\overset{..}{L}_{P},.....,L_{F},%
\overset{.}{L}_{F},\overset{..}{L}_{F},.....a\right) .  \label{eqn53}
\end{equation}

Also, the IR cut-off depends on other parameters such as the Hubble
parameter, the Ricci scalar, and their derivatives. Though, can be
transformed to either $L_{P}$ and their derivatives or $L_{F}$ and their
derivatives. The above-mentioned cut-off might be selected to be equivalent
to a general covariant gravity model

\begin{equation}
S=\int d^{4}\sqrt{-g}F\left( R,R_{\mu\nu}R^{\mu\nu},R_{\mu\nu\rho\sigma
}R^{\mu\nu\rho\sigma},\square R,\square^{-1}R,\nabla_{\mu}R\nabla^{\mu
}R,...\right) .  \label{eqn54}
\end{equation}

In the following, using the above expressions and with the help of the
generalized cut-off, we will show that the HDE of the present work has
direct equivalence to the generalized HDE model.

From Eq. (\ref{eqn24}) we get

\begin{equation}
\omega_{\Lambda}=-1+\frac{\left( 1+z\right) }{3}\frac{d}{dz}\left[ \ln\left(
\alpha H^{2}+\beta\overset{.}{H}\right) \right] ,  \label{eqn55}
\end{equation}
where, we used $dt=\frac{dz}{\overset{.}{z}}$, and $H$ is given by Eq. (\ref%
{eqn30}). The comparison of Eq. (\ref{eqn19}) with Eq. (\ref{eqn20}) and
using Eq. (\ref{eqn52}) immediately conduct to the equivalence holographic
cut-off (in terms of $L_{P}$ and its derivatives or in terms of $L_{F}$ and
its derivatives) corresponds to the HDE as

\begin{align}
\frac{3C^{2}}{k^{2}L_{R}^{2}} & =3\left\{ \alpha\left( \frac{\overset{.}{L}%
_{P}}{L_{P}}-\frac{1}{L_{P}}\right) ^{2}+\beta\left( \frac {\overset{..}{L}%
_{P}}{L_{P}}-\frac{\overset{.}{L}_{P}^{2}}{L_{P}^{2}}+\frac{\overset{.}{L}%
_{P}}{L_{P}^{2}}\right) \right\} ,  \label{eqn56} \\
& =3\left\{ \alpha\left( \frac{\overset{.}{L}_{F}}{L_{F}}+\frac{1}{L_{F}}%
\right) ^{2}+\beta\left( \frac{\overset{..}{L}_{F}}{L_{F}}-\frac {\overset{.}%
{L}_{F}^{2}}{L_{F}^{2}}-\frac{\overset{.}{L}_{F}}{L_{F}^{2}}\right) \right\}
.  \notag
\end{align}

The equation of state parameter can be derived from the conservation
equation corresponds to the HDE density $\rho_{HDE}$

\begin{equation}
\omega_{HDE}^{\left( RD\right) }=-1-\frac{\overset{.}{\rho}_{HDE}}{%
3H\rho_{HDE}}=-1+\frac{2}{3HL_{R}}\frac{dL_{R}}{dt},  \label{eqn57}
\end{equation}
where $L_{R}$ is given by Eq. (\ref{eqn56}). Hence, we conclude that $%
\omega_{HDE}^{\left( RD\right) }$ is equivalent to $\omega_{\Lambda}$ as
derived in Eq. (\ref{eqn55}).

\section{Conclusion}

In this paper, we have studied a model of HDE with a homogeneous and
anisotropic Universe of Bianchi type-I in the framework of $f(G)$ gravity.
In order to find exact solutions of the field equations, we assume that the
deceleration parameter (DP) varies with cosmic time. We discussed some
physical and geometric quantities of the model and got the following results:

\begin{itemize}
\item The deceleration parameter contains two phases in the Universe, the
initial deceleration phase and the current acceleration phase. The value of
the transition redshift for our model is $z_{tr}=0.62$, this value
corresponds to the observational data.

\item The scalar expansion and the shear scalar\ diverge at $t\rightarrow0$
and they become finite when $t\rightarrow\infty$. The anisotropic parameter\
remains constant throughout cosmic evolution, which indicates that our model
is puvely anisotropic from the initial era of the Universe to the final era.

\item From the evolution of the equation of state (EoS) parameter, we can
observe that in the primitive Universe $-1<\omega_{\Lambda}<0$, it indicates
the quintessential model and in the current Universe, $\omega_{\Lambda}$\
tends to $-1$, that is, the $\Lambda CDM$ model. Thus, the current value of
the equation of state parameter for our model is in good agreement with
recent observations.

\item The value of the density parameter is large in the first era of the
Universe, but it started to approach $\Omega\sim1$ in the last era of the
Universe, which leads our model to predict a flat Universe at a later time,
as recent astronomical observations indicate.

\item The cosmic jerk parameter is positive throughout the evolution of the
Universe and tends towards $1$ at late times.
\end{itemize}

In the literature, it is known that in the Standard Model, the cosmological
constant is the mechanism responsible for the current cosmic acceleration,
that is, the dominated phase of DE, which has negative pressure. In $f(G)$
gravity, responsible for this, are the additional terms of $f(G)$ next to
the scalar curvature $R$.

\textbf{Acknowledgments} We are very much grateful to the editor and
anonymous referee for illuminating suggestions that have significantly
improved our article.

\end{document}